\documentclass[useAMS,usenatbib]{mnras}
\usepackage{amsmath}
\usepackage{amssymb}
\usepackage{graphicx}
\usepackage{hyperref}

\usepackage{color}
\usepackage{pdflscape}

\newcommand{\oii}{[O\,{\sc II}]}

\newcommand{\oiii}{[O\,{\sc III}]}
\newcommand{\bpass}{{\sc BPASS}}

\title[Properties of z$\sim$5 LBAs]{Physical Properties of Local Star-Forming Analogues to $z\sim5$ Lyman Break Galaxies}

\author[Greis et al.]{Stephanie M. L. Greis$^{1}$\thanks{E-mail:
s.m.l.greis@warwick.ac.uk}, Elizabeth R. Stanway$^{1}$, Luke J. M.
Davies$^{2}$\newauthor \& Andrew J. Levan$^{1}$\\
$^{1}$Department of Physics, University of Warwick, Gibbet Hill Road, Coventry, CV4 7AL, UK\\
$^{2}$ICRAR, The University of Western Australia, 35 Stirling Highway, Crawley, WA 6009, Australia}
\begin{document}

\date{Accepted yyyy Month dd. Received yyyy Month dd; in original form yyyy Month dd}

\pagerange{\pageref{firstpage}--\pageref{lastpage}} \pubyear{2015}

\maketitle

\label{firstpage}

\begin{abstract}
Intense, compact, star-forming galaxies are rare in the local Universe
but ubiquitous at high redshift. We interpret the 0.1-22\,$\mu$m
spectral energy distributions (SED) of a sample of 180 galaxies at
$0.05<z<0.25$ selected for extremely high surface densities of
inferred star formation in the ultraviolet. By comparison with
well-established stellar population synthesis models we find that our
sample comprises young ($\sim$ 60 - 400 Myrs), moderate mass ($\sim
6\times 10^9$\,M$_{\odot}$) star-forming galaxies with little dust
extinction (mean stellar continuum extinction $E_\mathrm{cont}(B-V)
\sim 0.1$) and find star formation rates of a few tens of Solar masses
per year. We use our inferred masses to determine a mean specific star
formation rate for this sample of $\sim 10^{-9}$ yr$^{-1}$, and
compare this to the specific star formation rates in distant Lyman
break galaxies (LBGs), and in other low redshift populations. We
conclude that our sample's characteristics overlap significantly with
those of the $z\sim5$ LBG population, making ours the first local
analogue population well tuned to match those high redshift
galaxies. We consider implications for the origin and evolution of
early galaxies.
\end{abstract}


\section{Introduction}
Massive nearby galaxies, including the one we currently inhabit,
started their lives as low-mass, low-metallicity
star-forming galaxies in the early Universe. Many of
these smaller progenitors likely passed through a phase in which they
appeared as Lyman break galaxies, named after a distinctive
drop in their observed flux shortwards of their rest-frame Lyman-limit
at 912\,\AA\ caused by intervening clouds of neutral hydrogen.   These
galaxies are intensely star-forming systems, and are thought to be
comparable to the population whose ultraviolet flux caused the
reionization of the Universe before $z\sim6$ \citep[see
  e.g.][]{Harford2003, Rhoads2013,Bunker2010}. LBGs
are also primary sources for our understanding of the formation and
evolution of galaxies \citep[e.g.][]{StanwayDavies2014}. As well as being
relatively straightforward to select in observations, they are thought
to be the progenitors of ellipticals \citep{Douglas2007} and
spheroidal components of spiral galaxies \citep{Verma2007} in the
current epoch.

An estimate of the object's redshift is found by identifying the
redshifted Lyman break wavelength as lying between a pair of
photometric filters.  \citet{Steidel1996} developed the drop-out
technique using \textit{UGR} filters, thereby favouring the discovery
of $z\sim3-4$ LBGs, which drop out in the \textit{U}-band. Since then
many studies have been undertaken on both $z\sim3$ LBGs and their
local analogues. Direct observations of $z\sim3$ LBGs reveal their
rest-frame optical properties \citep[e.g.][]{Shapley2001}, luminosity
functions \citep[e.g.][]{vanderBurg2010}, UV-optical SEDs
\citep[e.g.][]{Papovich2001} and ultraviolet properties
\citep[e.g.][]{Rafelski2009,Ly2011,Boutsia2014}. Selected
 local galaxies, including highly ultraviolet-luminous starbursts
at $z\sim0.1-0.2$ \citep[UVLGs,][]{Heckman2005, Heckman2011} and the
extreme optical emission line galaxies \citep[EELS, EELGs or `Green
  Peas'][]{Cardamone2009, Amorin2015}, have been proposed as good
local laboratories in which to study the physics of the $z\sim3$
Universe. Detailed studies can be undertaken on such sources that
would be challenging on the LBGs themselves
\citep[e.g. spatially-resolved measurements of the gas
  kinematics, ][]{Goncalves2010}.

Colour-selected LBGs have been identified to
date at redshifts $1<z<10$, with studies of their bulk properties 
including \citet{Oteo2013} for $z\sim1$,
\citet{Hathi2013} and \citet{2012ApJ...745...96H} for $z\sim 1 - 3$,
\citet{Mosleh2012} and \citet{2015ApJ...810...71F}for $z\sim 1 - 7$, \citet{Huang2015} for lensed
$z\sim 6 - 10$ objects, and \citet{Bouwens2015} and
\citet{Ellis2013} for $z\ga9$ candidates). However, the detailed study
of high redshift LBGs, particularly at $z>5$ is complicated by their
small projected size and faint apparent magnitudes, pushing current
observations to their technical limits.  The interpretation of LBGs can therefore be
greatly improved by studying a more nearby sample of galaxies with
comparable star formation properties and physical sizes which act as
Lyman break analogues (LBAs).

In this paper we extend the pilot sample identified by
\citet{StanwayDavies2014} to define an analogue population of 180
candidate objects with $0.05<z<0.25$.  This sample selects on the two
key observable differences between $z\sim3$ and 5: ultraviolet
properties and physical size.  In this paper, we aim to answer the
question of whether our sample of nearby UV-selected galaxies is
indeed analogous to high redshift ($z\sim5$) LBGs and
to determined the specific star formation rates and galaxy
construction timescales for these sources. To answer these questions,
we undertake spectral energy distribution (SED) fitting of the far-UV
to near-infrared magnitudes of the objects, and determine their
masses, star-formation rates (SFRs), metallicities, dust contents, and
ages.

In section \ref{sec:Sample_Selection} of this paper we present the LBA
candidates sample, selected from SDSS DR7 and GALEX DR6, and
supplemented with spectroscopic data from the enhanced catalogue
provided by the MPA-JHU collaboration. The wavelength coverage is
extended to the infrared (using WISE and 2MASS data) in section
\ref{sec:2MASS and WISE photometry}. In section \ref{sec:SED Fitting}
we give an overview of the SED fitting procedure and stellar
population spectral synthesis codes used to determine the galaxies'
properties. We present the results of the SED fitting in section
\ref{sec:Results}, before presenting the inferred properties of our
sample in \ref{sec:inferred}. In section \ref{sec:LBAs?} we determine
the suitability of our sample as analogues to $z\sim5$ galaxies. We
present our conclusions in section \ref{sec:conclusions}.

Throughout this paper, magnitudes are given in the AB system. We adopt
a standard $\Lambda$CDM cosmology with
$H_0=70\,\mathrm{km\,s^{-1}\,Mpc^{-1}}$, $\Omega_M=0.3$ and
$\Omega_{\Lambda}=0.7$.

\section{Sample Selection}
\label{sec:Sample_Selection}

\subsection{Previous Work}
While analogue samples for $z\sim3$ Lyman break galaxies have been
established by \citet{Heckman2005} and \citet{Hoopes2007},
\citet{StanwayDavies2014} argue that they are an unsuitable analogue
populations for higher redshift ($z\sim5$) LBGs, as the redshift
evolution of Lyman break galaxies must be taken into account when
determining the appropriate selection criteria for their local
analogues.

LBGs at $z\sim5$ are less
luminous (see \citet{Bouwens2007}), younger and less massive (see
\citet{Verma2007}, \citet{Oesch2013}), and have lower metallicities
\citep{Douglas2010} than $z\sim3$ LBGs. The star formation rate
density within LBGs increases by a factor of $\sim 4-5$ from $z\sim5$ to
$z\sim3$ \citep{vanderBurg2010}. They are also significantly more
compact than their $z\sim3$ counterparts \citep{Wilkins2011,Mosleh2012}, which has implications for
the physical processes within the galaxies. The compactness modifies
star formation since a higher UV-photon density causes higher dust and
intergalactic medium temperatures, than would be present in more distributed star-forming 
regions, which thus affects the collapse of molecular clouds
into stars, the ionization of the intergalactic medium and potentially
the mode of star formation itself \citep[see
  e.g.][]{Stanway2014}. By comparing the star formation densities of the
\citet{Hoopes2007} sample with those of typical values for LBGs at
higher redshifts, \citet{StanwayDavies2014} indicate that only $\sim
3\%$ of the \citet{Hoopes2007} sample of UVLGs have a star formation
density comparable to that observed in distant LBGs. This discrepancy only becomes exacerbated as the Lyman break
samples are pushed to higher redshifts.  

In order to establish our sample, we modify the pilot sample
established by \citet{StanwayDavies2014}.  which was constrained to
declinations of $<-8^o$ to facilitate follow-up from Southern
telescopes. We lift this constraint, considering the full SDSS-GALEX
overlap region.  

\subsection{UV and Optical Selection}
Potential candidates are identified from data release six (DR6) of the
publicly available \textit{GALEX}
survey\footnote{http://galex.stsci.edu/GR6/} \citep{GALEX_Martin2005}
and DR7 of the Sloan Digital Sky Survey
\citep[SDSS,][]{Abazajian2009}, providing ultraviolet and optical data
respectively.  Candidates are selected such that their UV colours
satisfy $-0.5<FUV-NUV<0.5$ or $-0.5<FUV-r<1.0$, where \textit{FUV} and
\textit{NUV} correspond to the observed frame \textit{GALEX} far- and
near-ultraviolet bands at $\sim 1500$\AA\ and $\sim
2300$\AA\ respectively, while \textit{r} indicates the \textit{SDSS}
red band at $\sim6200$\AA . We set this restriction to ensure that the
rest-frame UV slope is close to flat, indicative of a recently formed
population of young stars with ages $<200\mathrm{Myrs}$. 

In addition to colour constraints, the luminosities of the potential
analogue galaxies are matched to those of high redshift
LBGs. \citet{Bouwens2007} find that the absolute magnitude of galaxies
becomes 0.7 mag brighter between $z\sim7$ and $z\sim4$, primarily due to
the hierarchical coalescence and merging of galaxies; we thus select
candidate galaxies such that their FUV absolute magnitudes are
equivalent to those of the existing $z>5$ LBG populations, by
requiring that $L_{UV}=0.1-5L^{\star}_{z=6}$ where
$M^{\star}_{UV}=-20.24$ at $z\sim6$ \citep{Bouwens2007}. Neither the
luminosities of the high redshift population nor those of this
analogue sample is corrected for dust extinction, as we discuss later.
In order to select comparable star formation densities to the
high redshift LBG population, candidates should have a projected
half-light radius $<2\mathrm{kpc}$. This is not always possible to
determine reliably in ground-based SDSS imaging. We allow for
unresolved sources, so that objects are selected which subtend
$<1.2''$ or $r_{1/2}<3.5$\,kpc in the galaxy's rest-frame where
this can be measured \citep[see][]{StanwayDavies2014}. It should be
noted that these constraints are applied using optical ($g$ and $r$ bands) rather than
ultraviolet images as available \textit{GALEX} images do not offer sufficient
resolution.

We require the candidates to have SDSS spectroscopy, which both
identifies the precise redshift of the galaxies and confirms the
source of their UV luminosity as most likely arising from star
formation instead of active galactic nucleus (AGN) activity. We
exclude galaxies with an AGN component identified by the SDSS
spectroscopic pipeline, since AGN are known to be rare in the distant
galaxy population; see \citet{Douglas2007} for $z\sim5$ or
\citet{Nandra2005} for $z\sim3$ observations. Additionally, we perform
an initial, by eye, assessment of the reliability of these sources to
exclude sources with clearly inaccurate photometry or multicomponent
sources in which only a small region satisfies our criteria. All
galaxies in the sample have very low mean Galactic foreground
reddening \citep{SF2011} ranging between $0.007\leq E(B-V)_{SFD}\leq
0.088$, with a mean extinction of 0.025. We adjust their photometry
for this using the Milky Way extinction law of \citet{Allen1976}.

\begin{figure}
\centering
\includegraphics[width=0.99\linewidth]{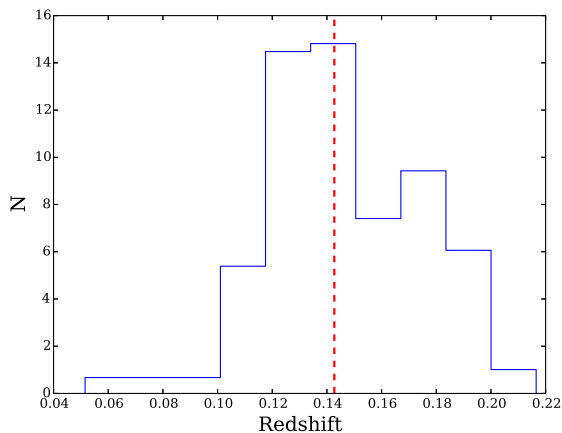}
\caption{Distribution over redshift $z$ of the sample of 180 candidate
  galaxies. The median redshift of $\sim$0.14 is indicated by the dashed
  vertical line.}
\label{fig:redshift_dist}
\end{figure}

The redshift distribution of our resulting sample of 180 candidate
galaxies at $0.05<z<0.25$ is shown in
Fig. \ref{fig:redshift_dist}.

\section{Infrared photometry}
\label{sec:2MASS and WISE photometry}
We determine photometry for the sample from the Widefield Infrared
Space Explorer \citep[WISE,][]{Wright2010} and 2-Micron All Sky Survey
\citep[2MASS,][]{Skrutskie2006} images, such that we have photometric
data spanning a combined wavelength range of approximately 1500 \AA\,
to 22$\mu$m.  For consistency we determine either measurements or
limits on the galaxies' magnitudes from the infrared imaging using
fixed aperture photometry at the SDSS locations and the recommended
aperture sizes appropriate to point sources. 
The sources were selected to have no nearby neighbours in the optical
or ultraviolet, and so did not require model-dependent deblending. However
in 10\% of the sample 2MASS photometry at the source location was unreliable
due to high local noise or close and blended neighbours.
Approximately a third of the candidate objects are individually
undetected above a $2\sigma$ detection limit in one or more infrared
bands, where this limit is determined locally in each band.

\section{SED Fitting}
\label{sec:SED Fitting}
A galaxy's spectral energy distribution can, in principle, provide
detailed information about its properties, including stellar mass,
star formation rate, metallicity, and dust content. The flux in
different wavebands is compared to synthetic spectra created from a
library of either empirical or stellar population synthesis (SPS)
model templates. These models comprise the integrated spectra of
artificial stellar populations, the shape and magnitude of which
depend on the star formation history, metallicity and initial mass
function (IMF), which themselves shape distributions in stellar age
and mass. As a result, the choice of SPS model can affect the derived
parameters.  Here we compare results using the
\citet[M05,][]{Maraston2005} SPS models against templates created using the
Binary Population and Spectral Synthesis SPS code \citep[\bpass,
][]{EldridgeStanway2009,EldridgeStanway2012}.

\paragraph*{Maraston models:}
As described in \citet{Maraston2005}, the Maraston models generate
composite stellar populations (CSP) with a Salpeter IMF and star
formation rates which decline exponentially with time such that,
$\dot{M}\propto e^{-t/\tau}$, where $\tau$ is the \textit{e}-folding
timescale.  These are built by combining simple single-age stellar
population models which have been calibrated against globular cluster
data for which ages and element abundances are independently known, so
that various generations of stars can be modelled
\citep{Maraston2005}. The parameters of the physical inputs
in the models, such as convection, mass loss, and mixing, which could
not be derived from first principles, are thus fixed by observations.

\begin{figure}
\includegraphics[width=\linewidth]{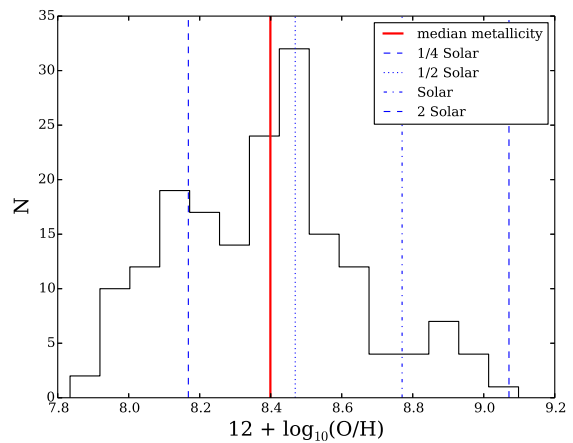}
\caption{The oxygen abundance distribution of our sample as
  calibrated from strong emission line ratios in SDSS spectroscopy,
  using the line diagnostic of \citet{2016Ap&SS.361...61D}. Dashed
  vertical lines indicate 0.25, 0.5, 1 and 2 times the local Galactic
  concordance value of 12+log(O/H)=8.77 which has been proposed as a
  better calibration than bulk Solar abundance. Solid vertical line
  indicates the sample median.}
\label{fig:R23vsOratio}
\end{figure}

For our SED fitting procedure, we
allowed timescales $\tau$, of 100\,Myr, 500\,Myr and 1\,Gyr. We
constrain the metallicity of our input models to 0.5\,Solar which
is a good match for the metallicity derived from nebular emission in
these sources, as derived from the calibration of \citet{2016Ap&SS.361...61D} and shown in
Fig.~\ref{fig:R23vsOratio}.

\paragraph*{BPASS models:}
The Binary Population and Spectral Synthesis code, or
\bpass\footnote{See http://bpass.auckland.ac.nz}, was developed to
address the effects of massive star evolution on the SED of galaxies
\citep[][Eldridge et al in
  prep]{EldridgeStanway2009,EldridgeStanway2012}. While in young
stellar populations a galaxy's spectrum is dominated by the radiation
emitted by massive hot stars, a more aged population shows a spectrum
which is strongly influenced by the evolution of moderately massive
stars that live for longer than the most massive population. This
evolution is often modified by processes such as angular momentum
transfer, and mass loss and gain due to a binary companion, allowing
evolved secondary stars to extend their highly luminous phase, and to
boost the population of rapidly rotating H-depleted Wolf-Rayet stars
\citep{Stanway2014}.  The standard \bpass\ distribution provides model
spectra for instantaneous starbursts. 
From these, we construct  exponentially declining star formation history models,
to match those available in the Maraston models. 

We use v2.0 of \bpass\ (Eldridge et al. in prep) and again select the
0.5\,$Z_\odot$ models and an initial mass function with a typical
Salpeter slope of $-2.35$ for stars between $0.5$ and $100M_{\odot}$,
and a slope of $-1.3$ for stellar masses between $0.1$ and $0.5
M_{\odot}$.

The stellar flux output in the \bpass\ model is processed to estimate
the nebular emission component using the radiative transfer code {\sc
  Cloudy} \citep{Ferland1998}.  We select an electron density of
$10^2$\,cm$^{-3}$, a covering fraction of 1, and a spherically
symmetric gas distribution with inner radius of 1 pc. These are
appropriate for a H\,{\sc II} region, as discussed in
\citet{EldridgeStanway2012} and \citet{Stanway2014}.

\paragraph*{Dust modelling:}\label{sec:dust-modelling}
The template spectra are modified using the \citet{Calzetti2000} dust
extinction law, which was empirically derived for local
infrared-luminous galaxies with active star formation.  In the
Calzetti law the extinction in front of nebular line emission regions
is higher than that on the stellar continuum such that
$E_\mathrm{cont}(B-V)$= $0.44\times\,E_\mathrm{line}(B-V)$. For
broadband photometry, the emission is dominated by stellar
continuum. 

At each model age, we modify the synthetic SED by a reddening curve
corresponding to colour excesses in the continuum,
$E_\mathrm{cont}(B-V)$, between 0.0 and 0.6 mags (in steps of 0.05).
We do not constrain the dust using the observed nebular emission since
it is likely that the nebular emission regions are significantly
smaller than the galaxy as a whole
\citep{2009ApJ...691..465F,2009ApJ...704L..98S}, and so the dust
measured by the Balmer decrement may well not be representative of the
stellar continuum emission.

\paragraph*{Fitting Procedure:}\label{sec:fitting_procedure}
A typical source in our sample is well detected in the optical and
ultraviolet, but often weakly detected or undetected in the
near-infrared. Their mid-infrared (WISE) properties vary
significantly. Given the number of parameters required both to build a
stellar population, and to be inferred from the resulting template,
any SED-fitting procedure can be subject to substantial uncertainty
and degenerate solutions. The number of extracted parameters can be
comparable to the number of input data-points.

Given that the relatively new BPASS models are not integrated into
most existing fitting codes, and in order to retain physical insight
into the input parameters, we construct our own SED fitting code. This
makes use of the well-established $\chi^2$ statistic, which is
appropriate for assessing goodness of fit in samples dominated by
Poisson noise, and for which the range of $\chi^2$ values representing
a 1\,$\sigma$ uncertainty band on any given parameter is determined by
a distribution dependent on the number of free parameters. The
uncertainty range quoted on numerical values is the range of models
considered with a $\chi^2$ value with a 68 per cent or better probability
of being consistent with the best fit model. We note that an
alternative would involve marginalising over the Bayesian priors for
this sample, however since the priors are poorly constrained, and the
data is sparse, this is unlikely to produce a stronger or more
reliable constraint in this circumstance.

We apply this method to fit the observational data over the spectral
range 0.12-2.16\,$\mu$m, i.e. from \textit{GALEX} FUV to
{2MASS} $K_S$ band inclusive. Where sources are undetected at
the 2\,$\sigma$ level in a given band, we use the 1\,$\sigma$ flux in
that band, and assign an uncertainty equal to the flux. We note that
at wavelengths longwards of $K_S$ the effects of dust extinction are
negligible and thus excluding the relatively shallow WISE bands does
not impact the inferred extinction.  Similarly, the reprocessing of
extincted ultraviolet emission by thermal dust does not produce
emission shortwards of 3\,microns and so no dust emission component is
required for modelling the stellar population.  We thus exclude the
\textit{WISE} bands from the fitting procedure since these are most
strongly dependent on the dust properties adopted rather than the
input stellar spectra. However we evaluate the consistency of the
\textit{WISE} data with our best fit model spectra in section
\ref{sec:wisefit}.

In order to evaluate the effectiveness and reliability of our spectral
fit we also make comparisons to two additional fitting methods. We
make use of the enhanced spectroscopic catalogue publicly available
from the MPA-JHU collaboration\footnote{based on DR7,
  http://www.mpa-garching.mpg.de/SDSS/DR7/} \citep{Brinchmann2004},
containing SDSS photometric and spectroscopic data as well as inferred
physical properties. These were derived from a narrower wavelength
range than we consider, but make additional use of spectroscopic
constraints. We find that a small number (3) of our objects are not
included in the MPA-JHU catalogues, but this has no significant effect
on the comparisons presented in later sections.  We also consider a
parallel analysis using the Maraston models and the {\sc CIGALE}
fitting algorithm
\citep{2005MNRAS.360.1413B,2014ASPC..485..347R}. This incorporates
stellar, nebular, and dust extinction components which we select to
range over the same values as our code. It also includes dust emission
and performs a marginalisation analysis to determine the best
fit. Configuring the {\sc CIGALE} dust emission component requires
three additional parameters and thus the number of constrained inputs
is comparable to the number of photometric data points on a typical
source. Nonetheless, throughout we confirm that this method produces
very similar results to those of our independent fitting code. We also
calculate fits extended either to the $K_S$ band (as is the case in
our own fitting) or to 22\,$\mu$m band. We confirm that there is no
significant difference in the observed stellar and dust extinction
components if the WISE bands are included in the fit, although several
additional parameters are required to do this.

\section{SED Properties of our Sample} 
\label{sec:Results}

\subsection{Masses}
\label{subsec:Masses}
The best-fitting mass is one of the most robust outputs of an SED
fitting procedure since it depends primarily on the optical/NIR
normalisation of the spectrum (i.e. the galaxy luminosity at its known
redshift) rather than the details of the spectral shape. However,
there is modest dependence on the synthetic stellar population input,
both in terms of that initial normalisation and since the mass to
light ratio at any given wavelength depends on the stellar population.

\begin{figure}
\includegraphics[width=\linewidth]{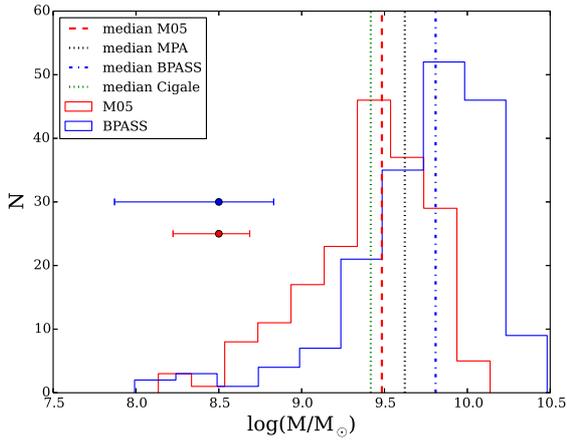}
\caption{The distribution of derived stellar masses, comparing the M05
  and \bpass\ models. The median fitted values are indicated by
  vertical dashed lines, as are those independently found by MPA-JHU
  and by fitting the data with the {\sc CIGALE} code and M05 models. These
  all show good agreement. Representative uncertainties on individual
  masses are indicated by points and associated ranges. The
  uncertainty on the median is substantially smaller.}
\label{fig:MedianMassStd_MPA_BPASS}
\end{figure}

The median stellar masses, M$_\ast$, of our sample were consistent between input
templates, and found to be log(M$_\ast$/M$_{\odot})=9.80\pm0.42$ for the standard
 \bpass\ model and log(M$_\ast$/M$_{\odot})=9.31\pm0.34$ for the M05 model.
In figure \ref{fig:MedianMassStd_MPA_BPASS}
 we compare the distributions of masses. As expected, the
derived distributions are broadly similar between different input SPS models,
with no catastrophic disagreements between them. The \bpass\ model
produces slightly higher masses than the equivalent M05 model.

The independent galaxy template fitting procedure performed for star-forming 
galaxies by the MPA-JHU collaboration \citep{Brinchmann2004}
shows good agreement with the SED model fits for our objects, as does
a parallel fitting procedure using the M05 models and {\sc CIGALE}. In the
following sections, we select the \bpass\ model results for further
analysis given the more physically motivated stellar population synthesis 
incorporated in these, and the generally weak dependence on input model
in best-fit mass.

\subsection{Ages}
\label{subsec:Ages}
\begin{figure}
\includegraphics[width=\linewidth]{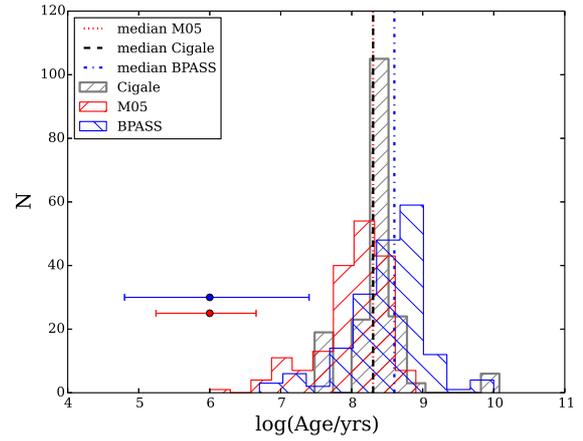}
\caption{Distribution of the stellar population ages of the target
  galaxies found using SED fitting with the M05 and \bpass\ 
  models. Indicative uncertainty ranges on individual galaxies are indicated by points.}
\label{fig:Age_M05_BPASStau}
\end{figure}

For most galaxies the age of the stellar population in an SED is most strongly
constrained by the strength of the spectral break at around
4000\AA\ in the rest-frame. Our selection criteria required the break
to be modest, with near-flat ultraviolet to optical colours, selecting
young starbursts.  The population as a whole shows a
relatively narrow distribution of best fit ages as figure \ref{fig:Age_M05_BPASStau} illustrates.
The uncertainty on individual galaxies (i.e. the range yielding a $\chi^2$ statistic with a
probability within 68 per cent of the best fit) is somewhat larger and typical values are indicated
on the figure.

The ages found using the M05 models were lower than the ones
found from the \bpass\ models, with median ages of
$\log\mathrm{(age/yr)}=7.78\pm0.49$ and
$\log\mathrm{(age/yr)}=8.60\pm0.52$ respectively. The higher age found
using \bpass\ $\tau$ models is consistent with the more massive stellar
populations found in section \ref{subsec:Masses}.  

\subsection{Dust Extinction}
\begin{figure}
\includegraphics[width=\linewidth]{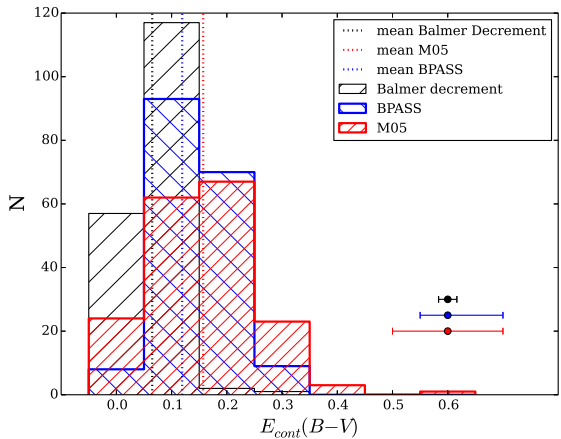}
\caption{The distribution of $E_\mathrm{cont}(B-V)$ values obtained
  from model SED fitting (blue for \bpass, red for M05), and those
  calculated from the observed H$_{\alpha}$/H$_{\beta}$ flux ratios
  (assuming case B recombination with a temperature of 10$^4$ K and an
  electron density of $n_e$= 10$^2$ cm$^{-3}$, multiplied by 0.44 to
  recover equivalent continuum extinction; black). }
\label{fig:MeanDust_BalmerBPASS}
\end{figure}

In addition to the foreground extinction arising from line of sight
through the Milky Way, we also fit for an internal dust extinction
component for each object, applied in the objects' rest-frame.  The
best-fitting $E_\mathrm{cont}(B-V)$ values are in close agreement
regardless of SPS templates adopted, giving mean $E_\mathrm{cont}(B-V)$ values of
0.16$\pm 0.10$ for the M05 templates and 0.12$\pm 0.07$ for the
\bpass\ models, as shown in figure \ref{fig:MeanDust_BalmerBPASS}.

We also calculate an estimate of dust extinction from the ratio of
recombination line strengths in the objects' spectra. For this we
assume an intrinsic Balmer decrement for case B recombination, at a
temperature of 10$^4$ K and an electron density of $n_e$= 10$^2$
cm$^{-3}$, such that H$_{\alpha}$/H$_{\beta}$=2.86. For comparison
with the stellar continuum derived extinctions, we adjust this
$E_\mathrm{line}(B-V)$ value by a factor of 0.44. As figure
\ref{fig:MeanDust_BalmerBPASS} shows, the resultant
$E_\mathrm{cont}(B-V)$ value distribution shows good agreement with
those calculated from model SED fits.

\section{Inferred Properties} 
\label{sec:inferred}

\subsection{Excitation Measurements}
\label{subsec:ExcitationMeasurements}
\begin{figure}
\includegraphics[width=\linewidth]{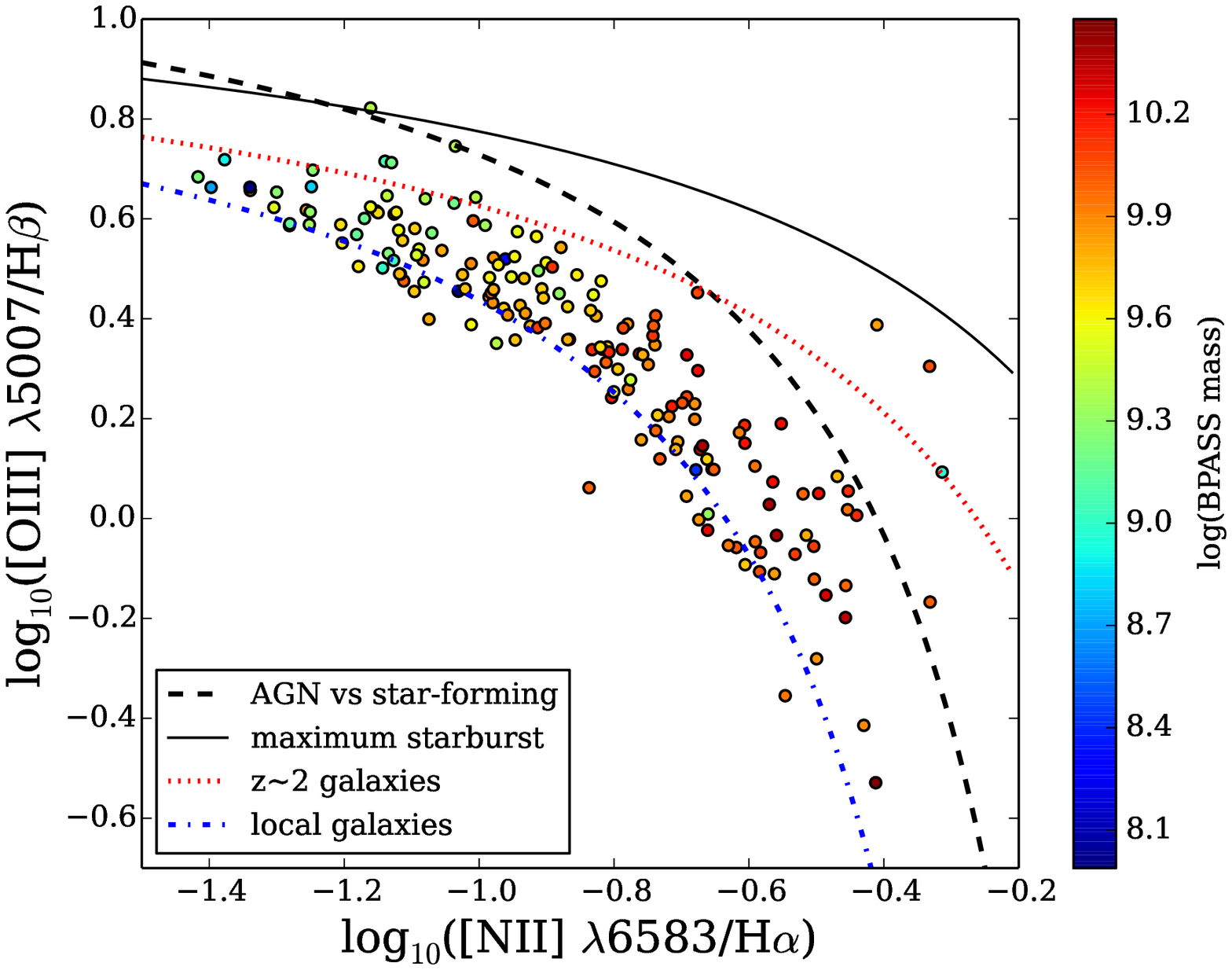}
\caption{BPT diagram for our sample, with colour coding according to
  the masses determined by SED fitting. The blue dot-dashed line
  indicates the average value for local galaxies \citep{Steidel2014},
  while the red dotted line shows the average of redshift $z\sim2$
  star-forming galaxies \citep{Steidel2014}. The solid black line
  indicates the proposed maximal starburst \citep{Kewley2001}, and the black
  dashed line shows the standard criterion used to separate AGN from
  star-forming activity \citep{Kauffmann2003}. Only a few targets fall
  above the dashed line, indicating that they might include an AGN
  component.}
\label{fig:BPT}
\end{figure}

The BPT diagram \citep{Baldwin1981} indicates the origin of the
ionizing radiation heating the nebular gas in a galaxy on the basis of its \oiii, H$\beta$,
[N\,{\sc II}] and H$\alpha$ flux ratios. This allows the classification of
the ionizing spectrum into star-forming or AGN-driven.  Using the BPT
diagram as a diagnostic for our sample (Fig. \ref{fig:BPT}) we
determine that the majority of sources fall below the proposed
cut-off separating star formation and AGN activity \citep[black dashed
  line,][]{Kauffmann2003}, thus confirming that these are indeed
star-forming galaxies and verifying the effectiveness of the SDSS spectroscopic
analysis pipeline. Four objects fall into the
`composite' region, indicating that these sources may contain components
of both star formation and AGN activity. While it is noteworthy that
our sample typically lies well above the median relation seen in local
galaxies, none of the objects in our
sample lie above the limit determined by \citet[][]{Kewley2001} for
a `maximal' starburst.

\begin{figure}
\includegraphics[width=\linewidth]{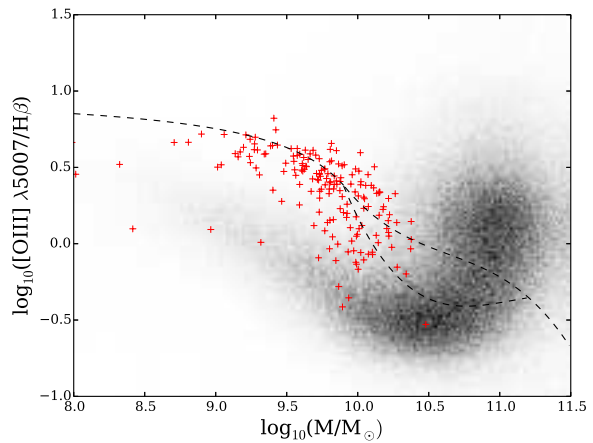}
\caption{Mass-excitation diagram for our sample (red crosses) and SDSS
  galaxies (underlying grey distribution).   The galaxies in our sample appear to be irradiated by a
  significantly higher ionizing potential than other local galaxies of
  comparable mass. The dashed lines indicate proposed
  classifications between AGN (above line), composite (bordered
  region), and star formation activity (below line) according to
  \citeauthor{Juneau2011} As in the BPT diagram, the majority of our
  objects are found in the star-forming region.}
\label{fig:OIII_Hbeta}
\end{figure}

The mass-excitation (MEx) diagnostic developed by \citet{Juneau2011}
provides an alternative indication of the AGN contribution within a
galaxy, similar to the BPT diagnostic but with the separation between
star-forming galaxies and AGN-hosts enhanced by the effects of the
mass-metallicity relation for local sources. Fig. \ref{fig:OIII_Hbeta} shows the
\oiii/H$\beta$ vs mass plane for the objects in our sample. Again, 
the majority of our objects trace the starburst galaxy region of the
parameter space, but, even more so than in the BPT diagram, we see an
offset between our population and the local SDSS sample from which it
was drawn, pushing our sample to straddle the border of AGN classification. 
Such a shift is likely indicative of a harder ionizing
flux, and associated higher ionization potential but not necessarily one arising from AGN
\citep[see][]{Stanway2014}.  This may indicate a difference in the galaxy
mass-metallicity relation between our sample and more typical galaxy
populations at the same redshift, with relatively little dependence of
metallicity on mass for these intense starbursts.

\subsection{SFRs}
\label{SFR}

\begin{figure}
\includegraphics[width=\linewidth]{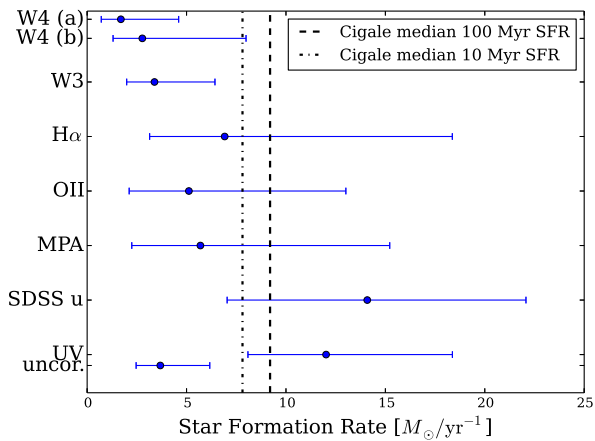}
\caption{The median, upper and lower quartile dust-corrected star
  formation rates (SFRs) of the sample. All fluxes were corrected for
  Milky Way dust attenuation. The \oii and H$\alpha$ fluxes were
  further corrected for the object-intrinsic dust extinction with
  $E_\mathrm{line}(B-V)$ values found using the (Milky Way
  dust-corrected) Balmer decrement, and the $U$-band and UV corrected
  using $E_\mathrm{cont}(B-V)$ determined by SED fitting. The UV SFR
  found without applying an intrinsic dust correction is shown
  underneath the corrected value. We show two calibrations for the W4
  band - that of \citet{2014ApJ...782...90C} as (a) and
  \citet{Lee2013} as (b). The SFRs determined by MPA-JHU are shown as
  a comparison. Sample medians derived using the {\sc CIGALE} parameter
  marginalisation code and the M05 stellar models, averaged over two
  different timescales, are indicated by vertical lines.}
\label{fig:SFR_comparison}
\end{figure}

In our SED fitting analysis we allowed star formation rate to vary as
this was constrained by the combination of SFR
timescale, mass, extinction and stellar population age. We note that
while it might have been possible to pre-constrain this using nebular
line emission, such constraints would have been dependent on dust
extinction, and star formation history assumed in the calibration
used. To explore the effect of such assumptions, and their
appropriateness for this sample, we determine the 
SFRs of our galaxies using a range of established SFR indicators, from UV to mid-infrared wavelengths.

\textit{UV:} We calculate SFRs from dust-corrected FUV
fluxes using the \citet{Madau1998} prescription at $1500$\,\AA\ for a
Salpeter IMF. For a stellar population with ongoing continuous star
formation dominated by young stars, the UV luminosity is a good tracer
of the stellar birth rate and independent of the star formation
history if the age $t\gg t_{MS}$, where t$_{MS} \leq 2\times10^7$ yrs
for late-O/early-B type stars \citep{Madau1998}. For comparison, we
also calculate SFRs based on the SDSS $u$-band photometry, after
correcting for dust, using the calibration of
\citet{2003ApJ...599..971H} - as figure \ref{fig:SFR_comparison}
shows, this is associated with a relatively large uncertainty, but is
comparable to the dust-corrected $FUV$-derived SFR.

\textit{H$\alpha$ and \oii:} We calculate the SFRs from H$\alpha$ and
\oii\ fluxes, using standard conversion factors (\citet{Kewley2004},
see also \citet{Kennicutt1998}). The SDSS fibre \oii\ and H$\alpha$
fluxes are not corrected for fibre losses since our targets are very
compact, but are corrected for nebular dust extinction as described
above.  Both strong line indicators are very sensitive to
instantaneous star formation, i.e. stellar ages $<3 - 10$ Myr, but
also sensitive to the ionization conditions and metallicity of the
nebular gas and so the applicability of these calibrations in this
sample has not been verified.  The UV and the spectral line indicators
agree within a factor of a few, and it can thus reasonably be assumed
that the galaxies are experiencing an ongoing starburst on timescales
of at least a few tens of Myr.

\textit{W3 and W4:} The SFRs inferred from the infrared bands were
found using the empirical \citet{Lee2013} prescription for WISE bands
W3 and W4. These SFR indicators were calibrated for a large sample of
local star-forming galaxies with SFR $>3$ M$_{\odot}$ yr$^{-1}$ and
mid-infrared data. Galaxies with apparent AGN activity were excluded
from the calibration. For W4, we also use the calibration of
\citet{2014ApJ...782...90C}, which is based on the local Galaxy and
Mass Assembly (GAMA) survey \citep{GAMA_DR2,2011MNRAS.413..971D}. The
median SFRs we obtain from W3 and W4 are 4.9 and 2.4 M$_{\odot}$
yr$^{-1}$ respectively (based on the Lee et al calibration). As
mentioned in section \ref{sec:wisefit}, the PAH emission region falls
into the W3 and W4 bands, and can dominate the infrared luminosity of
star-forming galaxies and hence this empirical calibration is somewhat
redshift-dependent. In addition, the emission from thermal dust
components has a relatively long timescale ($>$100\,Myr) to become
established after the onset of intense star formation and does not
always trace the youngest starbursts \citep{2015MNRAS.452..616D}. We
note that some of the W3 and W4 magnitudes in our sample were upper
limits, these have been included in this median with a nominal SFRs
equal to the 1$\sigma$ flux limits. As a result, the derived median
values may be skewed by these inferred limits on SFR.

The median, upper and lower quartile SFRs derived for
the entire sample are shown in Fig. \ref{fig:SFR_comparison}. The
different SFR indicators, as well as the independently determined
MPA-JHU results, are in good agreement, with the median values of each
indicator lying between $\sim 2.5$ and 14 M$_{\odot}$\,yr$^{-1}$,
albeit with large ranges. The highest SFRs are found from the
dust-corrected UV fluxes, but are associated with considerable
uncertainty on the dust extinction correction. We also show UV SFRs
without correction as they provide a lower limit on the SFR in the
objects. An additional constraint on SFR can be obtained from the SED
fitting procedure, although this depends on the star-forming
population dominating the stellar light. The star formation rates
derived from our \bpass\ models are comparable to the rest-UV
estimates with a median of $\sim$15\,M$_{\odot}$\,yr$^{-1}$. Those
derived from the M05 models (either using our fitting procedure or
{\sc CIGALE}) are slightly lower with a median around
$\sim$8\,M$_{\odot}$\,yr$^{-1}$. Interestingly, in both cases, the
average SFR over 100\,Myr timescales is higher than
the instantaneous SFR, suggesting that this is
decreasing over time for the median galaxy. Our sample's best-fitting
ages (as described in section \ref{subsec:Ages}) of a few tens to a
few hundreds Myrs, and the uncertain contribution of strong line
emission to the WISE bands, suggest that the H$\alpha$ and UV SFR indicators are most likely to provide accurate
descriptions of the objects, subject to uncertainties in extinction
and differences in emission timescale.

\subsection{Specific Star Formation Rates and Timescales}

\begin{figure}
\includegraphics[width=\linewidth]{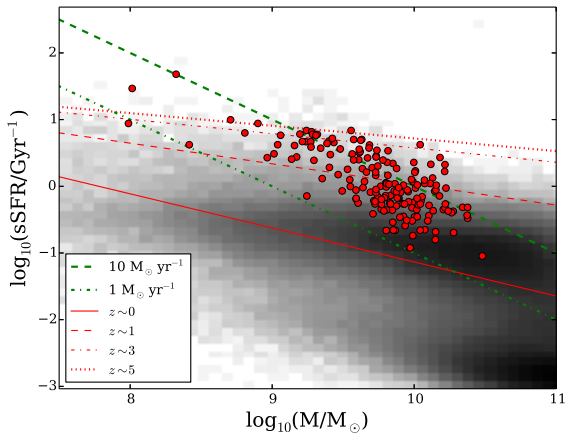}
\caption{Specific star formation rate (as log$_{10}$(sSFR/Gyr$^{-1}$))
  vs mass of our sample (red dots), using the H$\alpha$ SFR
  conversion. The green dashed and dot-dashed lines represent constant
  star formation rates of 10 M$_{\odot}$ yr$^{-1}$ and 1 M$_{\odot}$
  yr$^{-1}$ respectively. The red lines represent the star formation
  main sequence at different redshifts according to
  \citet{Leslie2016}. The underlying grey distribution gives the
  logarithm of the local SDSS galaxies' distribution. It is apparent
  that our objects lie significantly above the main sequence for local
  galaxies.}
\label{fig:sSFR_withSDSS}
\end{figure}

From the star formation rates calculated from H$\alpha$ in section
\ref{SFR} and the masses derived in section \ref{subsec:Masses}, we
calculate the specific star formation rates (sSFR = SFR/mass) and star
formation rate surface densities ($\Sigma_{SFR} =$ SFR/area).  The
galaxies in our samples have log(sSFR/yr$^{-1}$)$\sim
-9.00\pm0.47$. 
As shown in Fig. \ref{fig:sSFR_withSDSS}, these sSFRs indicate that our sample is more intensely
star-forming at a given stellar mass than more typical local
galaxies. Our galaxies lie significantly above the star-forming galaxy
main-sequence for $z\sim0$.  While our galaxies show a mild evolution
in sSFR with mass (consistent with our ultraviolet luminosity, and
hence SFR, selection criteria), we do not find any correlation of sSFR
with redshift over the narrow redshift range spanned by the sample.

\begin{figure}
\includegraphics[width=\linewidth]{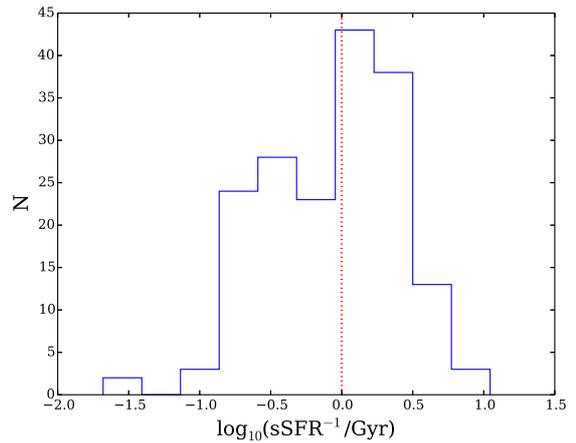}
\caption{The distribution of the inverse specific star formation
  rate. This corresponds to the mass doubling time scale of the
  galaxies. The median value of 1.0 Gyr is shown by the vertical red
  dotted line.}
\label{fig:inv_sSFR}
\end{figure}

The inverse of the specific star formation rate, sSFR$^{-1}$, provides
a measure of the time it would take to double the stellar mass of the
system assuming constant star formation at the present rate. In order
to calculate this mass doubling time scale, we use the star formation
rates derived from H$\alpha$. We find that the mass doubling time
scales of our sample have a median of 1.0 Gyr as shown in
Fig. \ref{fig:inv_sSFR}.

\subsection{Dust Emission}\label{sec:wisefit}
Our fit to the optical to near-infrared data is not sensitive to the
reprocessing of ultraviolet emission in the thermal infrared. However,
the observed extinction and inferred stellar continuum can be used to
predict the expected emission if there is no additional,
heavily-obscured star-forming or AGN component. To model the
re-emission of thermal photons at long wavelengths, we adopt the
energy-balance prescription of \citet{daCunha2008}. While this is
implemented in the {\sc MAGPHYS} code, it represents a simple formulation in
which the energy lost from UV-optical is re-emitted as a combination
of grey-body and polycyclic aromatic hydrocarbons (PAH) emission
components. We reproduce this in our own SED fitting code, using
identical black body parameters to those derived by
\citet{daCunha2008} as the {\sc MAGPHYS} defaults for star-forming
galaxies. We do not attempt to vary these parameters or fit the shape
of the dust emission curve, since two data points,
with typical signal to noise $<10$ on each, is insufficient to do
so. Instead, we simply consider constraints on normalisation for a
single possible emission spectrum. For the PAH emission component, we
used the average luminosity-weighted composite template of
\citet{Smith2007}, which is based on low resolution mid-infrared {\it
  Spitzer Space Telescope} spectroscopy for nearby AGN and
star-forming galaxies. The luminosity of intensely star-forming
galaxies in the mid-infrared ($\sim 3 - 25$ $\mu$m) is dominated by
strong emission features attributed to PAHs, with up to 20\% of the
infrared flux emitted in the strong transition lines alone.

\begin{figure}
\includegraphics[width=\linewidth]{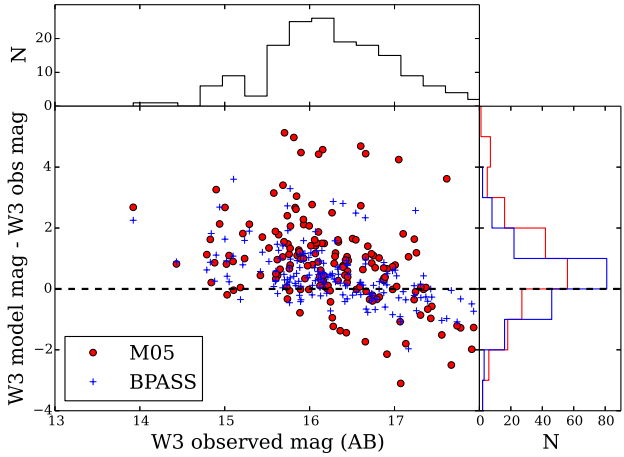}
\caption{Comparing the observed WISE 3 magnitude to the colour difference between the SED model and observations. The red dots indicate the M05 model results, while the \bpass\ model results are represented by blue crosses. The dashed line marks the magnitude line where model and observations coincide. For WISE 3, at 12$\mu$m, no significant difference can be seen between the two models.}
\label{fig:W3_obs_mod}
\end{figure}

\begin{figure}
\includegraphics[width=\linewidth]{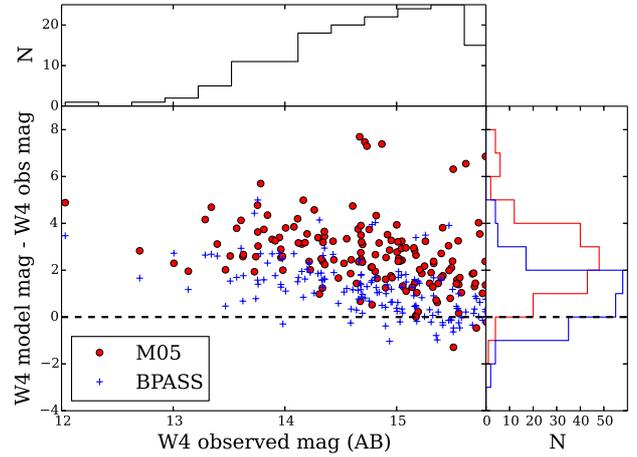}
\caption{Same as Fig. \ref{fig:W3_obs_mod} but for the WISE 4 band at $22 \mu$m. Both model sets underpredict mid-infrared emission. A clear offset between the M05 and \bpass\ model can be seen with the \bpass\ (blue crosses) model's magnitudes in better agreement to the observed ones.}
\label{fig:W4_obs_mod}
\end{figure}

Comparing the measured infrared fluxes (or upper limits) with our
best-fitting model SEDs, we find no significant discrepancy between
these in the near-infrared shortwards of the $K_S$ band. However, in
the PAH-dominated region of $\sim 3 - 25 \mu$m, we see an offset
between model and observation.  Figures \ref{fig:W3_obs_mod} and
\ref{fig:W4_obs_mod} show this offset for the WISE W3 and W4 bands
centred at 12 and 22 $\mu$m respectively. At 12 $\mu$m (observed) the
models marginally under-predict the galaxy flux, based on our fitted
extinction and assumed dust emission law. No significant difference in
behaviour is observed between the M05 and \bpass\ models.

In the W4 band, however, predicted magnitudes derived from the \bpass\ model are in
much better agreement with the observed values than those found using
the M05 model. This suggests that, given the combination of \bpass\
stellar models and \citeauthor{Calzetti2000} dust extinction, the
\citeauthor{Smith2007} composite luminosity-weighted average spectrum
provides a reasonable estimate of the PAH component. However, if the
M05 stellar population synthesis model better describes the
population, a steeper PAH model with increased flux at higher
wavelengths \citep[see Fig. 18 in][]{Smith2007} may be more
appropriate.

An extinction law which absorbs more flux
at short wavelengths would also lead to stronger IR dust emission and hence
stronger PAH features. Our use of the Calzetti law was motivated by
its description of intensely star-forming systems - exceeding the star
formation rate of the Milky Way or other sites, for which dust law
formulations exist, and more akin to the galaxies in our sample. While
it was originally developed for a small sample of UV and FIR-bright
galaxies, \citeauthor{Calzetti2000} also confirmed the applicability
of the dust reddening law on a larger sample of starburst galaxies,
and it is widely used for starbursts across a broad redshift range.

The thermal dust emission is modelled here as a series of single
temperature grey-bodies. A change in the dust grain size or
composition may modify these.  We note that fitting using the {\sc CIGALE}
code, which scales a dust emission curve derived from
\citet{2014ApJ...784...83D}, but otherwise has identical input stellar
populations to ours, overpredicts the median model flux by 40 per
cent. However, in deriving this, we had to assume three default dust
parameters to fit data essentially constrained by two points. We would
suggest that further investigation of the mid-infrared properties of
this or similar samples may be required.

\section{Suitability as Lyman break analogues}
\label{sec:LBAs?}

\subsection{Implications for $z\sim$5 LBGs}
Our sample was selected for their potential use as analogues to the
most distant galaxy populations. In table
\ref{table:ComparisonsOtherSamples} we summarise the inferred physical
properties of our sample, and also those derived for the high redshift
galaxy population. In many respects these are similar, with comparable
dust extinction, stellar mass, stellar age, metallicities, star
formation rates and sSFRs.

A small number of our larger, older galaxies may be inappropriate as
analogues for the distant galaxy population.  None of the objects in
our sample have masses greater than $5\times10^{10}$ M$_{\odot}$, and
132/180 objects satisfy a mass criterion of
M\,$<10^{10}$M$_{\odot}$. Further excluding all objects with ages
greater than one Gyr leaves 124/180 sources, while only accepting
those with ages $<10^{8.5}$ years, produces a young subsample of
78/180 galaxies (none of which show AGN features).

\begin{figure}
\includegraphics[width=\linewidth]{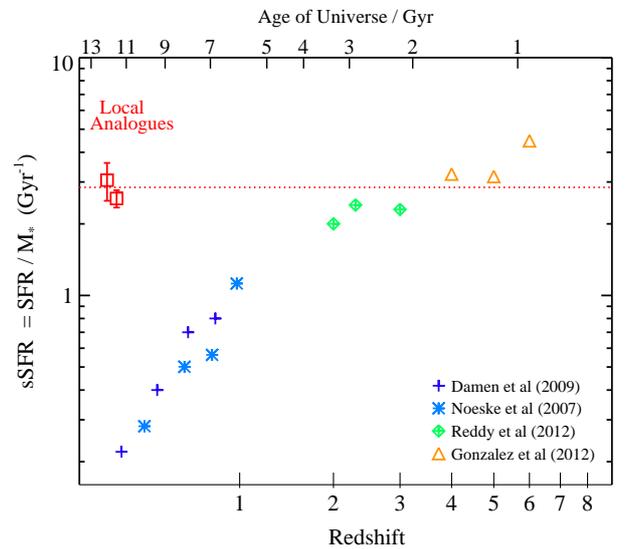}
\caption{The specific star formation rate (sSFR) of our sample (red
  squares) compared to other galaxy populations at different
  redshifts. We split our sample into two redshift subsets, at
  $0.05<z<0.15$ and $0.15<z<0.25$, and indicate the mean of the entire
  sample with a dotted line. The average sSFR of our sample is
  significantly higher than that of other local and low-redshift
  galaxies, but in very good agreement with the sSFRs found for
  $4<z<6$ LBGs.}
\label{fig:sSFR_redshift}
\end{figure}

In Fig. \ref{fig:sSFR_redshift} we consider the specific star
formation rate of our sample in comparison to more typical galaxy
populations as a function of redshift. We make use of
the high signal to noise H$\alpha$ measurement for SFR, and apply
a conservative correction, increasing the SFR by 25 per cent to
account for the more mature star formation component which would
be measured in the ultraviolet. We note that our UV-derived star
formation measures are larger than this corrected value, but associated
with large uncertainties on dust extinction. Thus we consider the
values in the figure as realistic lower limits on the sSFR. At low
redshifts our sample is very atypical of the bulk of the population. At high
redshifts, LBGs constitute the majority of the observable galaxy sample. It
is apparent that the sSFRs of our sample lie significantly above those
for more typical local galaxies (measured in H$\alpha$).  In this respect they are more akin to those
seen in $z\sim 4 - 6$ LBGs (as can also be seen in
Fig. \ref{fig:sSFR_withSDSS}), which are also an extreme,
ultraviolet-selected star-forming population.  The high sSFRs in this
sample also provide a possible explanation for the high excitation
parameters seen in them (see Fig. \ref{fig:OIII_Hbeta}) and poses
interesting questions. These include their escape fraction of
ionizing radiation, and its comparison to those inferred in
high redshift galaxies?

\begin{figure}
\includegraphics[width=\linewidth]{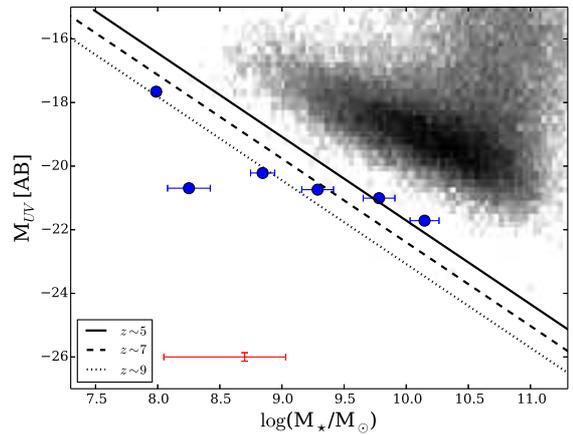}
\caption{The absolute (dust-corrected) 1500\,\AA\ $FUV$ magnitude vs
  mass distribution of our sample. The black solid line shows the
  theoretical prediction for $z\sim5$ galaxies by \citet{Dayal2014};
  lines for $z\sim7$ and 9 are also shown on the plot. The blue
  circles give the mean values of our sample, binned in 0.5 dex mass
  increments, and their standard errors. The M$_{UV}$ uncertainties
  are too small to be seen. The red error bars are indicative of the
  uncertainty on each individual datum contributing to a given
  bin. For comparison, the distribution for galaxies measured by the
  GAMA survey \citep{GAMA_DR2} at the same redshift as our sample is
  shown in grey.}
\label{fig:MUV_mass}
\end{figure}

We calculate the ultraviolet continuum mass to light ratio for our
sources, using masses determined via SED fitting and dust-corrected
1500\,\AA\ $FUV$ fluxes as shown in Fig. \ref{fig:MUV_mass}. We
compare this to the theoretical predictions of \citet[][dashed black
  line]{Dayal2014} for $z\sim5$ galaxies, and also with the
ultraviolet luminosities of more typical low redshift star-forming
galaxies selected from the GAMA survey
\citep[][greyscale]{GAMA_DR2}. For the latter, we apply the Calzetti
law and the extinction reported by the GAMA team to correct for dust
extinction. The Dayal models apply a prescription for supernova-driven
wind quenching of star formation based on a semi-empirical galaxy
evolution model. Our sources are offset from the local GAMA sample,
but in very good agreement for the predicted $z\sim5-9$ population,
suggesting that the properties of this sample may be driven by the
same evolutionary pressures and feedback constraints that apply in the
distant Universe.

If the galaxies comprising our sample
were redshifted to $z\sim 5$, corresponding to an age of the Universe
of $\sim1.15$ Gyr, their mean mass doubling time scale would place
their formation redshift at $z\sim 22$. However, this assumes constant
star formation over 1\,Gyr, which would seem unlikely. The median age of the
dominant stellar populations as derived from fitting M05 models would
indicate that they formed as late as $z\sim 5.2$, while the \bpass\ model
would place their formation redshift at $z\sim 7$ (based on the
population averages). The latter is in very good agreement with the
formation redshifts inferred for $z\sim 5$ LBGs, which
\citet{Lehnert2007} found to be $z\sim 6 - 7$.

\subsection{Comparison to Other Galaxy Populations}
\label{sec:Comparison_otherPops}

A number of galaxy populations have now been proposed as local
analogues to high redshift LBGs. While these share a common property -
the presence of intense star formation - they vary significantly in
their selection criteria and observed properties. In Table \ref{table:ComparisonsOtherSamples} 
we also contrast our sample with some of the other suggested local
analogue populations.

The most widely-used analogue sample is a selection of local ($z<0.3$)
UV-luminous ($L_{FUV}>2\times 10^{10} L_{\odot}$) galaxies (UVLGs)
chosen to overlap the luminosity range of $z\sim3$ LBGs
\citep{Heckman2005}.  The selection criteria for the original
\citet{Heckman2005} sample and ours are very similar. They aim to
reproduce the properties of similar galaxy populations, albeit
populations at different epochs in the evolution of the Universe. As a
result the galaxies presented in \citet{Heckman2005},
\citet{Hoopes2007}, and \citet{Goncalves2010} are significantly more
massive than our sample, reflecting the evolution of mass and
luminosity between $z\sim5$ and $z\sim3$ LBGs \citep[see
  e.g.][]{Verma2007}. We find that only one of our 180 objects
overlaps with the \citet{Hoopes2007} sample.

Similarly, blue compact 
dwarfs \citep[BCDs,][]{Zwicky1965} are characterised by compact and gas-rich regions of
high star formation. However, unlike our sample, BCDs are representative of the extremely low end of
galaxy luminosity ($M_B = -18$ mag), mass, and metallicity
functions. They are among the most metal-poor galaxies in the local
Universe, with some objects having metallicities below a tenth Solar,
below that inferred for our sample or for the high redshift population
at $z\sim5$. Thus, while both our sample and the BCD sample can be
characterised as compact star-forming galaxies, it is likely that the
BCD sample provides a better match for galaxy populations at still
earlier times ($z>8$), when both typical metallicity and typical
luminosity are expected to be rather lower than at $z\sim5$. None of
our sample would be classified as BCDs.

The `Green peas' or extreme emission line galaxies (EELGs) are a
sample of local ($0.112<z<0.360$) SDSS galaxies with strong nebular
emission lines, particularly the [O III] $\lambda 5007$\,\AA\ line in
the SDSS $r$-band giving rise to a green appearance
\citep{Cardamone2009,Amorin2015}.  However, there are significant
differences between our sample and these extreme emission-line
sources. While most of the Peas/EELGs are identified as star-forming,
some objects fall on the AGN or `composite' regions of the BPT
diagram.  Their selection technique also differs in important respects
from the LBG selection. They are required to show strong optical
nebular features, particularly the \oiii\ line, and no constraint is
placed on the ultraviolet continuum. This is effectively a selection
on emission line equivalent width, more akin to Lyman-$\alpha$ emitter
(LAE) selection than typical Lyman-break selection techniques.

To demonstrate the complementarity of our LBA sample and the EELGs,
it is interesting to consider the Hydrogen recombination line strengths.
High redshift ($z\ga5$) galaxies are observed in the rest-frame
ultraviolet and thus usually characterised by their Lyman-$\alpha$
emission line. Both the Pea/EELG sample and our own sample of
LBAs are observed in the rest-frame optical and their
properties therefore most easily quantified in the Balmer series. A
comparison between the two is not entirely straightforward.
Lyman-$\alpha$ is resonantly scattered and its radiative transfer can
be complex. Nonetheless, we can infer a predicted rest-frame
Lyman-$\alpha$ equivalent width from the H$\alpha$ feature in the low
redshift samples. To do so we use stellar population synthesis
models to model the scaling from H$\alpha$ equivalent
width to Lyman-$\alpha$ at the typical age of each sample. We also
adjust the inferred Lyman-$\alpha$ emission to recover the predicted
emission in the presence of dust, assuming a typical extinction
$E_{cont}(B-V)=0.1$, and account for the higher extinction of the emission
lines relative to the stellar continuum according to the prescription
of \citet{Calzetti2000}. 

\begin{figure}
\includegraphics[width=\linewidth]{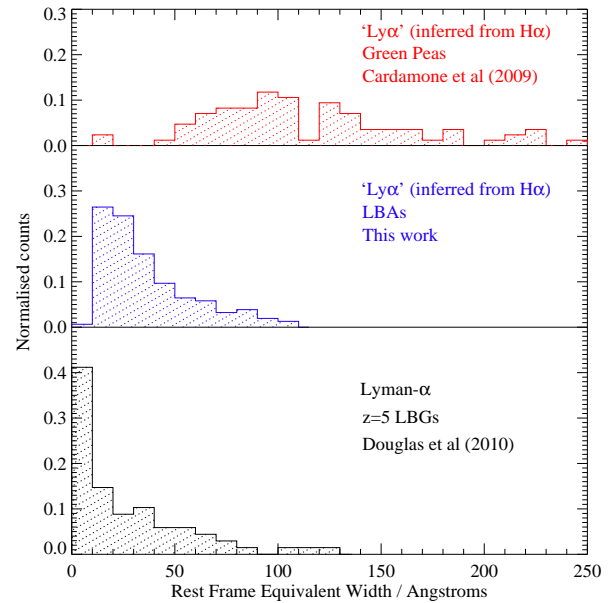}
\caption{The equivalent width distributions of Lyman-$\alpha$ in the
  $z\sim5$ Lyman break galaxy sample of \citet{Douglas2010}, in the
  Green Pea sample of \citet{Cardamone2009} and in our sample. In the
  latter two cases, a predicted Lyman-$\alpha$ equivalent width has
  been inferred from their Balmer line emission as discussed in
  section \ref{sec:Comparison_otherPops}.}
\label{fig:GP_ew}
\end{figure}

In figure \ref{fig:GP_ew}, we consider the comparison between the
Green Pea sample of \citet{Cardamone2009}, the $z=5$ sample of
spectroscopically-confirmed Lyman break galaxies from
\citet{Douglas2010} and the sample presented in this paper, based on
hydrogen line equivalent width. While neither analogue sample fully
reproduces the observed $z\sim5$ distribution, it is notable that only
22$\pm$4 per cent of the \citeauthor{Douglas2010} sample had a
rest-frame equivalent width exceeding 20\,\AA\ - quite unlike the
distribution seen in the EELG sample.
A significant
difference in equivalent width distribution is indicative of a
discrepancy in stellar population between samples since this is sensitive
to the ratio of instantaneous star formation rate and its longer term average.

\subsection{Future Work}
The study of this sample of analogue galaxies is still at an early
stage.  These galaxies provide an ideal laboratory to characterise the
expected behaviour of line ratios, metallicity indicators, gas
densities, and temperatures which will be directly probed at high
redshifts in the JWST/ELT era. Any discrepancy between their nebular
properties and those measured directly at high-$z$ may thus be
attributed to large scale intergalactic and radiation field properties
rather than the local intense starburst activity. Such physical
processes may underlie the evolution of the mass-metallicity relation.
Line emission analysis will also yield outflow speeds and geometry for
our objects, and the kinematics of different emission lines can
provide insights into the age of the outflows and accretion.

Measuring the sub-millimetre flux of this sample will allow us to
directly probe the obscured star formation in them, as well as putting
constraints on their dust mass, extinction and emission properties.
Exploring the gas masses of our objects will complement such studies,
enabling us to determine star formation efficiencies, and how the
fraction of gas in the galaxies has changed over
time. \citet{Davies2010} and others \cite[e.g.][]{Coppin2015} have
determined the molecular gas content of $z\sim5$ LBGs from CO line
emission.  Similar observations of our local analogue sample will
yield better constraints on the H$_2$ mass, and possibly detection of
CO lines not achievable even through stacking of high redshift images.

Finally, radio continuum and line observations of this sample will
provide an independent measurement of dust-obscured star formation
rate and, in combination with other SFR indicators, may be sensitive
to the age of the stellar population.   21\,cm line observations
will directly probe their atomic gas content, while similar
measurements at $z\sim5$  will not be possible before the
advent of the SKA.
 
\section{Summary and Conclusions}
\label{sec:conclusions}

In this paper we establish a population of galaxies which resembles
$z\sim5$ LBGs in their observed properties. Using colour cuts on SDSS
and GALEX archival data to identify objects which would fulfill the
drop-out criteria of LBGs, as well as UV-optical slopes indicative of
young stellar populations, we selected a sample of 180 $0.05<z<0.25$
galaxies.
We performed SED fitting on this sample using \citet{Maraston2005} and
\bpass\ \citep{EldridgeStanway2009,EldridgeStanway2012} stellar
population synthesis codes. 
Our analysis of best-fitting models 
allowed us to determine the objects' best-fitting ages, masses, and
dust content. In addition, we calculate star
formation rates, metallicities, and excitation measurements. Both observed
and inferred properties have been compared to those
determined for different galaxy populations, including $z\sim5$ LBGs.

Our main findings include the following: 
\begin{enumerate}
\item We determine a median stellar mass for our sample of $\log$(M$_\ast$/M$_{\odot}$)$\sim 9.80 \pm 0.42$, in agreement
  with the masses found for $z\sim 5$ LBGs. The median age of the
  sample is $\log$(age/yr)$=8.60\pm0.52$,
  indicating that, if our sample was redshifted to $z\sim5$, their
  formation redshifts would be $z\sim 6 - 7$. The $E_\mathrm{cont}(B-V)$ values,
  found using the \citet{Calzetti2000} starburst extinction law,
  showed little to moderate dust reddening with a median of 0.12$\pm0.07$.

\item Using observed flux measurements as well as spectroscopic data,
  we determine median star formation rates between $\sim 2.5$ and 14
  M$_{\odot}$ yr$^{-1}$, depending on SFR indicator, and mean specific
  star formation rates of $\sim 10^{-9}$ yr$^{-1}$.

\item Comparing these properties to those of $z\sim5$ LBGs in table
  \ref{table:ComparisonsOtherSamples}, we find good agreement. We
  therefore conclude that our sample can be used, with caution, as a
  local analogue population.
\end{enumerate}

\section*{Acknowledgements}
SMLG is funded by a research studentship from the UK Science and
Technology Facilities Council (STFC). ERS also acknowledges support
from STFC consolidated grant ST/L000733/1 and from the University of
Warwick's Research Development Fund.

This paper has made use of data from the Sloan Digital Sky Survey
(SDSS). The Sloan Digital Sky Survey (SDSS) is a joint project of The
University of Chicago, Fermilab, the Institute for Advanced Study, the
Japan Participation Group, The Johns Hopkins University, the Los
Alamos National Laboratory, the Max-Planck-Institute for Astronomy
(MPIA), the Max-Planck-Institute for Astrophysics (MPA), New Mexico
State University, University of Pittsburgh, Princeton University, the
United States Naval Observatory, and the University of
Washington. Funding for the project has been provided by the Alfred
P. Sloan Foundation, the Participating Institutions, the National
Aeronautics and Space Administration, the National Science Foundation,
the U.S. Department of Energy, the Japanese Monbukagakusho, and the
Max Planck Society. www.sdss.org is a winner of the Griffith
Observatory's Star Award.

This publication makes use of data products from the Two Micron All
Sky Survey, which is a joint project of the University of
Massachusetts and the Infrared Processing and Analysis
Center/California Institute of Technology, funded by the National
Aeronautics and Space Administration and the National Science
Foundation. This publication also makes use of data products from the
Wide-field Infrared Survey Explorer, which is a joint project of the
University of California, Los Angeles, and the Jet Propulsion
Laboratory/California Institute of Technology, funded by the National
Aeronautics and Space Administration.

We also make use of Ned Wright's very useful online cosmology
calculator \citep{2006PASP..118.1711W} and the {\sc TOPCAT} table operations software \citep{2005ASPC..347...29T}.

For data analysis, we make use of the {\sc IRAF} software
environmnet. {\sc IRAF} is distributed by the National Optical
Astronomy Observatory, which is operated by the Association of
Universities for Research in Astronomy (AURA) under a cooperative
agreement with the National Science Foundation.

\bsp
\label{lastpage}

\begin{landscape}
\begin{table}
\begin{center}
    \begin{tabular}{ | l | l | l | l | l | l | l | l |}
    \hline
    Sample & redshift & mass (median) & dust extinction & age & SFR & sSFR & metallicity\\
 \\
      & & log$_{10}$(M$_{\odot}$) & $E_\mathrm{cont}(B-V)$ & log$_{10}$(yrs) & M$_{\odot}$ yr$^{-1}$ & log(yr$^{-1}$) & Z$_{\odot}$ \\
 \hline \hline
    Our Sample$^a$ & 0.05 - 0.25 & $9.80\pm0.42$ & $\sim 0.12\pm0.07$ & $\sim 8.1-9.0$ & $\sim 14$ (UV) & $\sim -9.0\pm0.5$ & $\sim 0.5$ \\
 \hline \hline
     $z\sim5$ LBGs$^b$ & $\sim5$ & $8 - 11$ ($\sim 9$) & $\sim 0.2$ & $<8$ & $\sim$ few 100s & $\sim -8.7^f$ & $\sim 0.2$\\
 \hline
     LBAs$^c$ to $z\sim3$ & $0<z<0.3$ & $\sim 9.0$ to 10.9 & $A_{FUV}\sim 0 - 2$mag & $\sim9$ (doubling time) & few - 100 & -9.8 to -8 &  0$\sim 0.3 - 1.5$  \\
 \hline
     Green Peas$^d$ & $0.112<z<0.360$ & 8.5 - 10 & $\leq 0.25$ & $\sim 8$ (doubling time) & $\sim 10$ & up to -8 & $\sim 1$ \\
 \hline
     Blue Compact Dwarfs$^e$ & $z<0.01$ & 7-8 & low & $\sim$6-10$^g$ & $\sim 10^{-3} - 1$ & - & $\sim \frac{1}{10}$\\
 \hline
   \hline
    \end{tabular}
    \caption{b =\citet{Verma2007,Douglas2010,Yabe2009} from; c = from \citet{Heckman2005}, \citet{Hoopes2007}, \citet{Goncalves2010}; solar metallicity is taken as log[O/H] +12 $\sim$ 8.7; d = \citet{Cardamone2009}, e = \citet{Corbin2006}, \citet{Zhao2011}; f=\citet{Gonzalez2010}; g = the younger (1-10 Myr) population dominates the light while older (10 Gyr) stars dominate the mass}
    \label{table:ComparisonsOtherSamples}
\end{center}
\end{table}
\end{landscape}

\end{document}